# Security Analysis of Cloud Computing

Frederick R. Carlson
Saint Petersburg College
Saint Petersburg, Florida
352-586-2621

fcarlson@ieee.org

## ABSTRACT
This paper produces a baseline security analysis of the Cloud Computing Operational Environment in terms of threats, vulnerabilities and impacts. An analysis is conducted and the top three threats are identified with recommendations for practitioners. The conclusion of the analysis is that the most serious threats are non-technical and can be solved via management processes rather than technical countermeasures.

## Categories and Subject Descriptors
K.6.5 [**MANAGEMENT OF COMPUTING AND INFORMATION SYSTEMS**]: Security and Protection – *authentication, physical security, unauthorized access*

## 1. INTRODUCTION
Cloud computing provisioning companies are predicted to grow at a 20 percent compounded annual growth rate according to the technology consulting firm Gartner.[1] Currently this space has an estimated 59 billion dollar market size for public and hybrid cloud structures. There are many different implementations of cloud computing. The most popular are Infrastructure as a Service (IaaS) - renting of computing and storage; Platform as a Service (PaaS) - renting of remote platform hosting; and Software as a Service (SaaS) - renting of software services. The business case for this largely rests with organizations seeking to shift their Information Technology (IT) costs from Capital Expenditure Accounts (CapEx) to Operational Expenditure Accounts (OpEx).

One more set of terms about Cloud Computing Technology is important to understand the issues surrounding the cloud, namely the distinction between Public Cloud, Private Cloud and Hybrid Cloud, as follows:

1) Public Cloud - A public cloud is contained outside an organization. An example of this type of service is Amazon Web Services Elastic Cloud Computing offering, which generates an "instance" of a network, server or application interconnected by the public Internet.

2) Private Cloud - A private cloud is contained inside an organization. An example of this type of service is an Enterprise Virtualization of PCs using thin client technology where the instances of the PC are delivered to the user's desktop from a centralized server plant. The distinction that needs to be made is that the virtualized assets of the private cloud are behind some sort of organizational boundary from the public Internet, whereas the public cloud is not.

3) Hybrid Cloud (Part Public/Part Private) - Most cloud implementations are hybrid clouds. Even the most locked down private cloud likely has egress point(s) into the public Internet, if for no other reason than to access global infrastructure services such as Domain Name System (DNS) assets.

The major issue concerning organizations is the security of cloud offerings. The pooling of assets along with issues concerning virtualization are just a few of the rather new issues that have been introduced to the security community as this technology gets deployed. One of the major ramifications for this is that the Confidentiality, Integrity, Availability (CIA) model requires modification to include Multi-party Trust, Mutual Auditability and Usability to form a "CIAMAU" model when talking about securing cloud computing assets.[2] These additions to the security model are reserved for future work and do not affect the conclusion of this paper.

In 2010 and 2011, two of the major cloud computing services (Amazon and Google) had major outages.[3, 4] Neither one of these outages was due to a security issue, but these outages showed just how pervasive cloud computing has become as the outages gathered media coverage that included international attention. The issue with Google Documents was a change that exposed a memory management bug that only existed under very high loads. As organizations move to the cloud, most of the consequences of dysfunction will be economic and financial. There is a possibility of seeing health and safety issues regarding cloud-based systems that support utility and medical infrastructure. This possibility has potentially very severe consequences. In our analysis, the time required to fix the issue will be viewed as a major indicator of the impact of the event. In this paper, availability is weighed as the major impact and needed to fix the issue as a major metric.[5]

## 2. ASSETS TO BE ANALYZED
An asset is what needs to be protected from threats to cloud security. The types of assets considered in this analysis are computing and data elements in the cloud that affect one or more of the following: People, Activities and Operations, Information, Facilities and Equipment, and Materials.[6] The way these elements are affected by them is via a standard CIA Framework.[7] Table 1 maps these assets of value with the cloud vulnerabilities, and organizes them into three categories: People, Processes/Operations, and Technical.[8,9,10]

It is interesting to note that in the cloud computing space, most of the vulnerabilities affect the people, activities, operations, and informational assets of the organization. Only the physical security vulnerability affects all of the asset classes.

Criticality is a measure of consequences (high, medium or low) if an asset is lost or degraded. The greater the threat of survival or viability to its owners, those nearby, or to others who depend on the asset, the more critical it is.[11]

Table 1. Description of Assets of Value

| Assets of Value | Vulnerabilities |
|---|---|
| People, Activities and Operations, Information, Facilities and Equipment, and Materials | Physical Security |
| People, Activities and Operations, Information | Data Lockout |
| People, Activities and Operations, Information | Loss of Data |
| People, Activities and Operations, Information | Common Stack Vulnerabilities |
| People, Activities and Operations, Information | Execution Control |
| People, Activities and Operations, Information | Multiparty Cloud Config |
| People, Activities and Operations, Information | Authorization Issues |
| People, Activities and Operations, Information | Real Time Access Issues |
| People, Activities and Operations, Information | Identity Management |
| People, Activities and Operations, Information | Security of Virtual App |
| People, Activities and Operations, Information | Int and Ext Cloud App Interaction |
| People, Activities and Operations, Information | Trusted APIs |
| People, Information | Privacy of Data |
| | Legend |
| | People - Blue |
| | Processes/Ops - Red |
| | Tech - Green |

Criticality is a very challenging thing to measure in the cloud computing space as the assets are pooled. This creates transitive risks throughout the cloud framework that are simply not present in standard systems where there is a non-shared portfolio of assets. Criticality can really only be brought into play when a detailed list of instances is compiled regarding the assets an organization puts into a cloud environment. In this analysis, the terms Impact, Severity and Probability are used to achieve a similar result.

Cloud computing has the potential to create unique and novel cascading effects that will cause a multiplier effect on the asset that is attacked.[12] There are serious implications to these cascade effects when cloud computing becomes ubiquitous, particularly in utility systems that have moved Supervisory Control and Data Acquisition (SCADA) and SCADA-like processes to the cloud. These effects are critical to understand, but unfortunately are beyond the scope of this paper.[13]

Finally, vertically integrated effects need to be considered. The combination of vertically integrated effects and cascading effects can cause feedback loops in the attack that can drastically increase its effect.[14] These effects are also beyond the scope of this paper, but they should be noted, as they are a major concern.

## 3. VARIABLES IN THE MODEL

### 3.1 "CIA" Defined

Confidentiality is "Preserving authorized restrictions on information access and disclosure, including means for protecting personal privacy and proprietary information…" [44U.S.C.,Sec.3542].[15]

Integrity is "Guarding against improper information modification or destruction, and includes ensuring information non-repudiation and authenticity…" [44U.S.C.,Sec.3542].[16]

Availability is "Ensuring timely and reliable access to and use of information…" [44U.S.C.,Sec.3542][17]

In cloud computing, three additional areas are added to the standard Confidentiality, Integrity, Availability domains. The ones that are appended to CIA are Multi-party Trust, Mutual Auditability and Usability.[18]

### 3.2 Threats

An organization seeks to protect assets from threats. Cloud computing threat analysis reveals unique challenges. One reason for this is the difficulty of maintaining the chain of logic between assets, vulnerabilities, and threats. The difficultly arises from the common pool nature of cloud computing and the break between customers and service providers. As will be shown, a vulnerability that begins as a "People" vulnerability can transform into a threat that is more akin to a "Process" problem. The Saripalli & Walters paper deals with this issue by adding three more categories to the CIA model to reflect the transitive and shared nature of this issue. This paper will attempt to preserve the chain of logic with the CIA framework.[19]

This paper addresses five Technical Threats to Cloud Computing Assets. These threats are:

1) Management Interface Compromise (Technical) - Capturing vulnerabilities while using customer management front ends.[20] This threat is where a man-in-the-middle or some similar attack gets between the service provider and the management front end. Amazon Web Services has a feature where the user must download and authenticate with a public/private key pair to protect against this issue.

2) Intercepting Data in Transit (Technical) - Failure in transmissions security leads to data sniffing and man-in-the-middle attacks during transit.[21] This a more generalized man-in-the-middle attack than the Management Interface Compromise. In this type of attack, the actual production data going to and from the cloud is compromised.

3) Data Leaks Between Customer and Cloud Providers. (Technical)[22] - Intentional or unintentional "data leakage" between the customer and the cloud providers occurs. This issue is a symptom of a much more serious process issue of disconnects between the customer and the cloud provider(s).

4) Distributed Denial of Service (Technical)[23] – The attacker may use up all the metered resources of the customer.[24] The goal is to achieve "resource exhaustion" which is when the system is taxed more than it can handle.

5) Hypervisor Compromise (Technical)[25] – The core of the cloud has vulnerabilities. A hypervisor is the core of the virtualization technology that is at the heart of most cloud computing offerings. The integrity of hypervisor systems is a potentially serious threat.

This paper addresses five Operational/Process Threats to Cloud Computing Assets. These threats are:[26]

6) Isolation Failure (Processes) - Failure in separating storage, memory and routing causes Isolation Failure.[27] Isolation failure is an operational issue (and/or possibly a people issue) in virtually all cases. The interesting thing about this issue is that many other derivative threats and dysfunctions can arise from this issue. This is an issue that is relatively easy to resolve.

7) Insecure or Ineffective Deletion of Data (Processes) - Improper deletion of data creates this threat and is particularly relevant when moving cloud providers.[28] This issue is another process-oriented issue that can be handled by effective management.

8) Conflicts Between Customer Handling/Hardening Procedures and Cloud Environment (People).[29] - As this paper will show, this issue is the most pressing threat in the Cloud Computing Space.

9) Cloud Provider Malicious Insider (People) - Cloud provider employee maliciously alters, corrupts or locks out customer data to create this threat.

10) Resource Exhaustion (People) – The over- or under-provisioning of cloud resources due to mismanagement or attack creates this threat.[30]

## 3.3 Vulnerabilities

Vulnerability is a gap or weakness in security that can be exploited by threats to gain unauthorized access to an asset. As was shown in Table 1, this paper will categorize vulnerabilities in three ways - People, Processes/Operations and Technical. The vulnerability list from the paper written by Sengupta, Kaulgud & Sharma will be used, and will be categorized into these bins.[31] A quick overview of these vulnerabilities is in order. Sengupta, Kaulgud & Sharma see four distinct areas of vulnerabilities to the cloud environment. They are:

    1) Cloud Infrastructure, Platform and Hosted Code.

    2) Data, Data Integrity, Data Remanence, Data Privacy.

    3) Access - Authentication, Authorization, and Access Control (AAA), encrypted data communications, and user identity management.

    4) Regulation.[32]

All of these will be categorized except for number 4 (Regulation), which is an interesting and important topic, but, like cascade effects, is beyond the scope of this effort.

The three People-related vulnerabilities are 1) Physical Security of Data Center, 2) Data Lock Out in case of Provider Failure, and 3) Prevention of Data Loss.[33] Two of these: (Data Lock Out and Prevention of Data Loss), may also be seen as process errors. However, it is difficult to comprehend how a cloud provider could not have the processes in place to deal with these issues. In fact, if the management of the cloud provider does not have a solution in place to prevent these issues, it goes to a "people" issue at the management level.

The following seven vulnerabilities are grouped into the Operation/Processes Bin. All come from the Sengupta, Kaulgud & Sharma paper.[34]

1) Common Stack Vulnerability - This is where a virtualized instance of one customer somehow bleeds or affects another customer's instance.

2) Execution Control - This is where the customer has abstracted execution and the control is mostly in the hands of the provider.

3) Data Remanence - This is where the customer's data is somehow exposed after a move or a change in cloud providers.

4) Multiple Party Cloud Confidentiality - This is related to the Common Stack Vulnerability above, but is more specific to confidentiality concerns.

5) Authorization – This is ensuring that only authorized assets are available.

6) Access issues - As the provider is closer to this data, there is a concern that unauthorized access may occur.

7) Identity Management[35]

Last, these four vulnerabilities from Sengupta, Kaulgud & Sharma will be assigned to the Technical Bin.

1) Security of the Virtualized Application - This is a more technical reading of the Common Stack Vulnerability. Where common stack issues are really a process issue, this issue goes more into the technical set up of the hypervisor.

2) Internal and Cloud Based Application Interaction and Isolation – this is a lower level reading of "Execution Control" and "Multiple Party Cloud Confidentiality" from the prior section.

3) Trusted APIs - This issue goes to the cloud provider doing his due diligence and only using signed and current API instances.

4) Privacy of Data[36] - This issue goes to the ability of the cloud provider and/or user to comply with laws, norms and standards regarding private data such as Personally Identifiable Information (PII).

Then, ideally, potential Threats poised to exploit the Vulnerabilities stated should be addressed. For example, identity thieves could be a Threat poised to exploit the Vulnerability of poor training, and thereby cause a failure of Confidentiality. Notice the chain of logic between the Assets of Value, Vulnerability, and Threat, linked to (in this example) Confidentiality. The granularity in this chain of logic is important but this paper will focus on a threat-based analysis and leave the strict but necessary linkage between Vulnerability and Threat for future work.

## 3.4 Threat Impacts

Impacts will now be addressed. Table 2 shows the Threat Impact to Cloud Computing Assets. The Impact is stated as "High" or "Low" and is scored as a 2 for a High ranking and a 1 for a Low ranking. Because of the unique nature of pooled computing resources, the impact is rarely a Low ranking. As a general rule, the scoring tended to be Low if the threat could be addressed in a

| Table 2 – Threat Impact ||
| --- | --- |
| **Threats** | **Impact** |
|  | **High - 2, Low - 1** |
| Resource Exhaustion | 2 |
| Malicious Insider | 2 |
| Interception Data in Transit | 2 |
| Data Leaks | 2 |
| Distributed Denial of Service | 1 |
| Hypervisor Compromise | 2 |
| Isolation Failure | 2 |
| Improper Deletion of Data | 1 |
| Conflicts between customer procedures and cloud provider procedures | 2 |
| Insecure interface and APIs | 2 |
| Physical Theft | 2 |
|  |  |
|  |  |
|  |  |
| Legend |  |
| People - Blue |  |
| Processes/Ops - Red |  |
| Tech - Green |  |

rapid manner of time. Distributed denial of service is one example of a low threat level.

## 3.5 Risk Level

Risk results from threats exploiting vulnerabilities to obtain, damage or destroy assets. The Impact derived from Table 2, when combined with the credibility of Threat and degree of Vulnerability, yields a Risk Level (High, Medium, or Low). The Risk for each area (Confidentiality, Integrity, and Availability) must be stated. Shown in Table 3 (Risk Metrics and Calculation per Threat) is the Risk Level. The goal of this calculation is a "probability score" for each threat. This approach is distinctly utilitarian and is used to find the top three threats. The analysis is very simple and straightforward. For each threat, a "level of severity" is scored in terms of confidentiality, integrity and availability.

Scoring the "Low" threats with a value of 2, the "Medium" threats with a score of 4 and the "High" threats with a score of 8 begins the next phase. The paper then sums these three values to assess a "Severity Score". Then, the "Impact" value is factored in by multiplying the "Impact" by the "Severity Score" to obtain the "Total". At this point, the probability that this event will happen must be determined.

The next task is issuing a "Level of Probability" as "Low", "Medium" or "High" whilst assigning a value of 1, 2, or 4 respectfully. That score is then multiplied by the "Total" to get a score in the "Probability" field. This score, when ranked high to low, shows which are the most serious threats (High) and the least serious ones (Low). The definition of Risk (R) used is the Probability of the event (P) times its Severity (S) times its Impact (I) or $R = P * S * I$.

Table 3 - Risk Metrics and Calculation per Threat.

| Threat | Impact | Risk Level | | | | | |
|---|---|---|---|---|---|---|---|
| | High - 2, Low - 1 | Category Confidentiality - C Integrity - I Availability - A | Level of Severity (L=2, M=4, H=8) | Severity Score | Total (Severity x Impact) | Level of Probability (L=1,M=2, H=4) | Probability Score |
| Resource Exhaustion | 2 | C<br>I<br>A | 2<br>2<br>8 | 12 | 24 | 1 | 24 |
| Malicious Insider | 2 | C<br>I<br>A | 8<br>8<br>2 | 18 | 36 | 2 | 72 |
| Interception Data in Transit | 2 | C<br>I<br>A | 8<br>8<br>4 | 20 | 40 | 1 | 40 |
| Data Leaks | 2 | C<br>I<br>A | 8<br>8<br>2 | 18 | 36 | 1 | 36 |
| Distributed Denial of Service | 1 | C<br>I<br>A | 2<br>4<br>8 | 14 | 14 | 1 | 14 |
| Hypervisor Compromise | 2 | C<br>I<br>A | 4<br>4<br>4 | 12 | 24 | 1 | 24 |
| Isolation Failure | 2 | C<br>I<br>A | 8<br>4<br>2 | 14 | 28 | 1 | 28 |
| Improper Deletion of Data | 1 | C<br>I<br>A | 8<br>4<br>2 | 14 | 14 | 1 | 14 |
| Conflicts between customer procedures and cloud provider procedures | 2 | C<br>I<br>A | 4<br>8<br>2 | 14 | 28 | 4 | 112 |
| Insecure interface and APIs | 2 | C<br>I<br>A | 2<br>4<br>8 | 14 | 28 | 1 | 28 |
| Physical Theft | 2 | C<br>I<br>A | 8<br>8<br>8 | 24 | 48 | 2 | 96 |
| | | | | | | | |
| Legend | | | | | | | |
| People - Blue | | | | | | | |
| Processes/Ops - Red | | | | | | | |
| Tech - Green | | | | | | | |

## 4. RECOMMENDATION OF SECURITY CONTROLS

This paper will now endeavor to make intelligent, defensible and implementable recommendations. A protection, detection and correction scheme will be used to address these countermeasures. The only additional data is a summation section which rank orders the "Probability Score" and identifies the top three threats for the application of security controls.

The top threat to address is a process issue. It is the conflict between customer procedures and cloud provider procedures. This seemingly straightforward issue is more complex than meets the eye. As was stated earlier in the paper, the issue of shared context between the cloud service provider and the customer remains a serious and divisive issue. The second most severe issue is one of simple physical theft of the cloud computing assets. This has been the source of many of the data breaches, particularly in the health care sector. The third most serious issue is one of the malicious insider. It is interesting to note that none of these issues are particularly complex to understand.

## 5. COUNTERMEASURES

Table 4 (Countermeasures) shows the top three issues and the top three countermeasures. They are not complex issues and they are not complex solutions. Organizations would do well to implement these very basic processes.

Table 4 – Countermeasures

| Issue | Countermeasure |
|---|---|
| Conflicts between customer procedures and cloud provider procedures | Exhaustively researched and staffed Service Level Agreement Contracts |
| Physical Theft | Standard FISMA Physical Security Procedures |
| Malicious Insider | Standard FISMA Personnel Security Procedures |

## 6. CONCLUSION

Cloud computing is here to stay. The security issues around this technology are real. The actions that senior management must ensure happen before the organization moves to use this technology are as follows:

### 6.1 Service Level Agreements Are Critical

Organizations must have solid Service Level Agreements (SLA) with the Cloud Provider. These agreements should include what happens if the Cloud Provider goes insolvent and the terms of violation of the SLA should be extremely punitive. The web article "Seven Lessons to learn from Amazon's Outage" makes excellent points about this.[37] The major point made is that Amazon's cloud service was down for four days and they still did not technically breach their SLA. There is no way that an organization can ensure security unless the SLA contract is thought through completely.

### 6.2 Physical and Personnel Standards

Organizations must implement Federal Information Security Management Act (FISMA) standards in the area of Physical Security, both at the customer site and at the cloud provisioning sites. Physical access to the actual virtual machines or the storage means that all of the data of the organization will be considered compromised.

Management may see it as unusual that this cutting edge technology has such pedestrian issues as major concerns but management should not be fooled. These are expensive and complex domains and require relentless discipline to sustain. Addressing these issues give significant value for an organization to move to the cloud securely. It may well be that by moving on these issues first that the more complex issues will be much simpler to solve.

A contingency plan would be to partially transition to the cloud for low risk systems and test the technology. This plan has risks if cloud computing provides the competitive advantages that are predicted.

---

[1] (2010). *The influence of cloud in outsourcing*, 2010-2011. *Gartner Research (G00208940)*

[2] Saripalli, P., & Walters, B. (2010). *QUIRC: A Quantitative Impact and Risk Assessment Framework for Cloud Security*. Paper presented at 2010 IEEE 3rd International Conference on Cloud Computing, Miami, Florida. Retrieved from http://ieeexplore.ieee.org/xpl/freeabs_all.jsp?arnumber=5557981

[3] Warren, A. (2011, September 9). *What happened to Google Docs on Wednesday?* [Web log message]. Retrieved from http://googleenterprise.blogspot.com/2011/09/what-happened-wednesday.html

[4] Wainewright, P. (2011, April 24). *Seven lessons to learn from Amazon's outage* [Web log message]. Retrieved from: http://www.zdnet.com/blog/saas/seven-lessons-to-learn-from-amazons-outage/1296?tag=mantle_skin;content

[5] Moteff, J. The Library of Congress, Congressional Research Service (2004). *Risk management and critical infrastructure protection: assessing, integrating, and managing threats, vulnerabilities and consequences* Retrieved from http://ndu.blackboard.com/bbcswebdav/xid-469767_2

[6] Saripalli, P., & Walters, B. (2010). *QUIRC: A Quantitative Impact and Risk Assessment Framework for Cloud Security*. Paper presented at 2010 IEEE 3rd International Conference on Cloud Computing, Miami, Florida. Retrieved from http://ieeexplore.ieee.org/xpl/freeabs_all.jsp?arnumber=5557981

[7] Cloud Security Alliance, (2011). *Top threats to cloud computing v1.0* Retrieved from http://www.cloudsecurityalliance.org/topthreats/csathreats.v1.0.pdf